\documentclass[prb,showpacs,showemail,preprintnumbers,twocolumn,amsmath,amssymb,superscriptaddress]{revtex4-1}

\usepackage[T1]{fontenc}
\usepackage[applemac]{inputenc}
\usepackage[english]{babel}
\usepackage{graphicx}
\usepackage{amssymb}
\usepackage{amsmath}
\usepackage{overpic}
\usepackage{color}
\usepackage[T1]{fontenc}

\newcommand{\be}{\begin{equation}}
\newcommand{\ee}{\end{equation}}
\newcommand{\bea}{\begin{eqnarray}}
\newcommand{\eea}{\end{eqnarray}}

\def\m{\mu}

\def\th{\theta}

\def\s{\sigma}

\def\ra{\rightarrow}

\def\Ra{\Rightarrow}

\def\lb{\label}
\def\pref#1{(\ref{#1})}

\newcount\bozza \bozza=1
\ifnum\bozza=1
\newdimen\shift \shift=-2truecm
\def\lb#1{%
{\label{#1}\rlap{\kern\shift{$\scriptstyle#1$}}}}
\else\def\lb#1{\label{#1}} \fi

\begin{document}

\title{Non-linear $IV$ characteristics in two-dimensional superconductors: 
Berezinskii-Kosterlitz-Thouless physics vs inhomogeneity}

\author{G. Venditti}
\affiliation{ISC-CNR and Dep. of Physics, Sapienza University of Rome, P.le A. Moro 5, 00185 Rome, Italy}
\author{J. Biscaras}
\affiliation{Sorbonne Universit\'e, CNRS, MNHN, Institut de Min\'eralogie de Physique des Mat\'eriaux et de Cosmochimie, IMPMC, F-75005 Paris, France}
\author{S. Hurand}
\affiliation{Laboratoire de Physique et d'Etude des Mat\'eriaux, ESPCI Paris, PSL Research University, CNRS, 10 Rue Vauquelin - 75005 Paris, France}
\affiliation{Institute Pprime, UPR 3346 CNRS, Universit\'e de Poitiers, ISAE-ENSMA, BP 30179, 86962 Futuroscope-Chasseneuil Cedex, France}
\author{N. Bergeal}
\affiliation{Laboratoire de Physique et d'Etude des Mat\'eriaux, ESPCI Paris, PSL Research University, CNRS, 10 Rue Vauquelin - 75005 Paris, France}
\affiliation{Universit\'e Pierre and Marie Curie, Sorbonne-Universit\'es, 75005 Paris, France}
\author{J. Lesueur}
\affiliation{Laboratoire de Physique et d'Etude des Mat\'eriaux, ESPCI Paris, PSL Research University, CNRS, 10 Rue Vauquelin - 75005 Paris, France}
\affiliation{Universit\'e Pierre and Marie Curie, Sorbonne-Universit\'es, 75005 Paris, France}
\author{A. Dogra}
\affiliation{National Physical Laboratory, New Delhi, 110012, India}
\author{R. C. Budhani}
\affiliation{Department of Physics, Morgan State University, Baltimore, Maryland 21251, USA}
\author{Mintu Mondal}
\affiliation{School of Physical Sciences, Indian Association for the Cultivation of Science, Jadavpur, Kolkata 700032, India}
\affiliation{Tata Institute of Fundamental Research, Homi Bhabha Rd, Colaba, Mumbai 400005, India}
\author{John Jesudasan}
\affiliation{Tata Institute of Fundamental Research, Homi Bhabha Rd, Colaba, Mumbai 400005, India}
\author{Pratap Raychaudhuri}
\affiliation{Tata Institute of Fundamental Research, Homi Bhabha Rd, Colaba, Mumbai 400005, India}
\author{S. Caprara}
\affiliation{ISC-CNR and Dep. of Physics, Sapienza University of Rome, P.le A. Moro 5, 00185 Rome, Italy}
\author{L. Benfatto}
 \email{lara.benfatto@roma1.infn.it}
\affiliation{ISC-CNR and Dep. of Physics, Sapienza University of Rome, P.le A. Moro 5, 00185 Rome, Italy}
\date{\today}

\begin{abstract}
One of the hallmarks of the Berezinskii-Kosterlitz-Thouless (BKT) transition in two-dimensional superconductors is the 
universal jump of the superfluid density, that can be indirectly probed via the non-linear 
exponent of the current-voltage $IV$ characteristics. Here, we compare the experimental measurements of $IV$ 
characteristics in two cases, namely NbN thin films and SrTiO$_3$-based interfaces. While the former display 
a paradigmatic example of BKT-like non-linear effects, the latter do not seem to justify a BKT analysis. Rather, 
the observed $IV$ characteristics can be well reproduced theoretically by modelling the effect of mesoscopic 
inhomogeneity of the superconducting state. Our results offer an alternative perspective on the spontaneous 
fragmentation of the superconducting background in confined two-dimensional systems. 
\end{abstract}

\maketitle

The progress in material science has made nowadays available a wide class of systems with thickness 
ranging from a few nanometers down to the atomic-layer limit. The possibility to engineer these effectively 
two-dimensional (2D) materials in field-effect devices opens also the exciting possibility to tune 
their quantum-mechanical ground state by changing the electron density. In some remarkable cases, including 
transition-metal dichalcogenides \cite{reviewTMD}, SrTiO$_3$-based oxide interfaces \cite{reviewSTO}, such 
as LaAlO$_3$/SrTiO$_3$  and LaTiO$_3$/SrTiO$_3$ (LTO/STO), and the recently discovered twisted graphene 
\cite{herrera_nature18}, the ground state can be continuously tuned from metallic/insulating to superconducting 
(SC). How the reduced dimensionality influences both phases is still a largely 
open question, which challenges our basic understanding of the collective fluctuations in 2D systems. 

A particularly interesting issue about 2D SC materials regards the very nature of the SC transition, that  
is expected to belong to the same Berezinskii-Kosterlitz-Thouless (BKT) universality class of the 2D $XY$ 
model \cite{bkt,bkt1,bkt2}. This expectation holds in particular when the system is thin and 
dirty enough that the Pearl length exceeds the sample size and screening effects due to charged supercurrents 
can be neglected \cite{beasly_prl79}. The relevant excitations in this case are topological vortex-like 
configurations of the phase, and the energy scale is set by the superfluid stiffness 
$J_s=\hbar^2 n_s/4m= \hbar^2 c^2 d/16\pi e^2 \lambda^2$, 
where $n_s$ is the 2D superfluid density, $\lambda$ the penetration depth, and $d$ the film thickness. Within 
the BKT scenario, the transition to the normal state is driven by the thermal unbinding of vortex-antivortex pairs, 
that leads to specific signatures, the most striking being the discontinuous jump \cite{nelson} of $J_s$  
from a finite value right below $T_{BKT}$ to zero above it, with an 
universal ratio $J_s(T_{BKT}^-)/T_{BKT}=2/\pi$. This feature is in principle observable via direct measurements of 
$\lambda(T)$, or it can be inferred from the non-linear exponent of the $IV$ characteristics
\cite{hn79}, that is ruled by the breaking of vortex-antivortex pairs induced by 
a large enough current.

In practice, the experimental observation of the BKT transition in real systems is far from being straightforward. 
In clean thick films $J_s$ is much larger than the 
critical temperature, so that the temperature $T_{BKT}$ where $J_s(T_{BKT})\simeq T_{BKT}$ is  
indistinguishable from the $T_c$ at which pairing disappears. In few-nanometer thick films of conventional 
superconductors, like NbN or MoGe,  $n_s$ (and then $J_s$) is strongly suppressed by disorder 
\cite{epstein_prl81,epstein_prb83,fiory_prb83,lemberger_prl00,armitage_prb07,armitage_prb11,kamlapure_apl10,
mondal_bkt_prl11,yazdani_prl13,yong_prb13,ganguly_prb15}, making the BKT scale experimentally accessible.
A similar condition can be reached in STO-based interfaces, where an extremely fragile SC condensate was recently
reported \cite{bert_prb12,bergeal_natcomm18,caviglia_cm18}. However, in both cases the suppression of the stiffness 
comes along with an increasing {\em inhomogeneity} of the SC background, questioning the very applicability of the 
standard theoretical expectations based on the {\em clean} $XY$ model 
\cite{coura_prb05,benfatto_prb09,meir_epl10,mondal_bkt_prl11,meir_prl13,benfatto_review14,mirlin_prb15,maccari_prb17,maccari_cm18}. 
 In the case of thin films  of conventional superconductors the SC backgrounds fragments into islands 
with typical size of tens of nanometers 
\cite{sacepe_11,mondal_prl11,pratap_13,noat_prb13,roditchev_natphys14,leridon_prb16,brun_review17}, 
as indeed theoretically predicted when the phase-coherent SC state competes with the localization effects due to 
strong disorder \cite{trivedi_prb01,dubi_nat07,ioffe,nandini_natphys11,seibold_prl12,lemarie_prb13}. 
As a consequence, the superfluid-density jump is smeared out but is still observable either via the direct 
measurement of the inverse penetration depth 
\cite{lemberger_prl00,armitage_prb07,armitage_prb11,kamlapure_apl10,
mondal_bkt_prl11,yazdani_prl13,yong_prb13,ganguly_prb15}, 
or via the measurement of the exponent of the non-linear $IV$ characteristics near $T_c$ 
\cite{epstein_prl81,epstein_prb83,fiory_prb83}. On the other hand, for STO-based interfaces there has been 
increasing evidence that the SC background fragments in islands of larger size 
\cite{biscaras_natmat13,bid_prb16,jespersen_prb16,scopigno2016,caviglia_natcomm18,hurand_prb19}, explaining, e.g., the 
considerable broadening of the resistive transition as percolation via a network 
of SC puddles \cite{benfatto_prb09,caprara_prb11,caprara_sust15}. In this case the inhomogeneity can be triggered both by extrinsic effects, like domain structures in the STO substrate\cite{moler_natmat13,moler_prb16,kalisky_natmat17,kalisky_prb17}, and by intrinsic ones, like an electronic phase separation due either to the non-rigidity of the band structure at the 
interfacial potential well \cite{scopigno2016}, or to a strong density-dependent Rashba spin-orbit coupling 
\cite{caprara_prl12}.
For what concerns the BKT physics, the direct 
measurement of the $J_s$ is rather challenging, and few experimental reports exist so far 
\cite{bert_prb12,bergeal_natcomm18,caviglia_cm18}. As a consequence, the occurrence or not of a BKT-like transition has been 
usually inferred from the analysis of the $IV$ characteristics
\cite{triscone_science07,han_apl14,bid_prb16,caviglia_prb17}. 

\begin{figure}[htb]
\centering
\includegraphics[width=0.4\textwidth,angle=0]{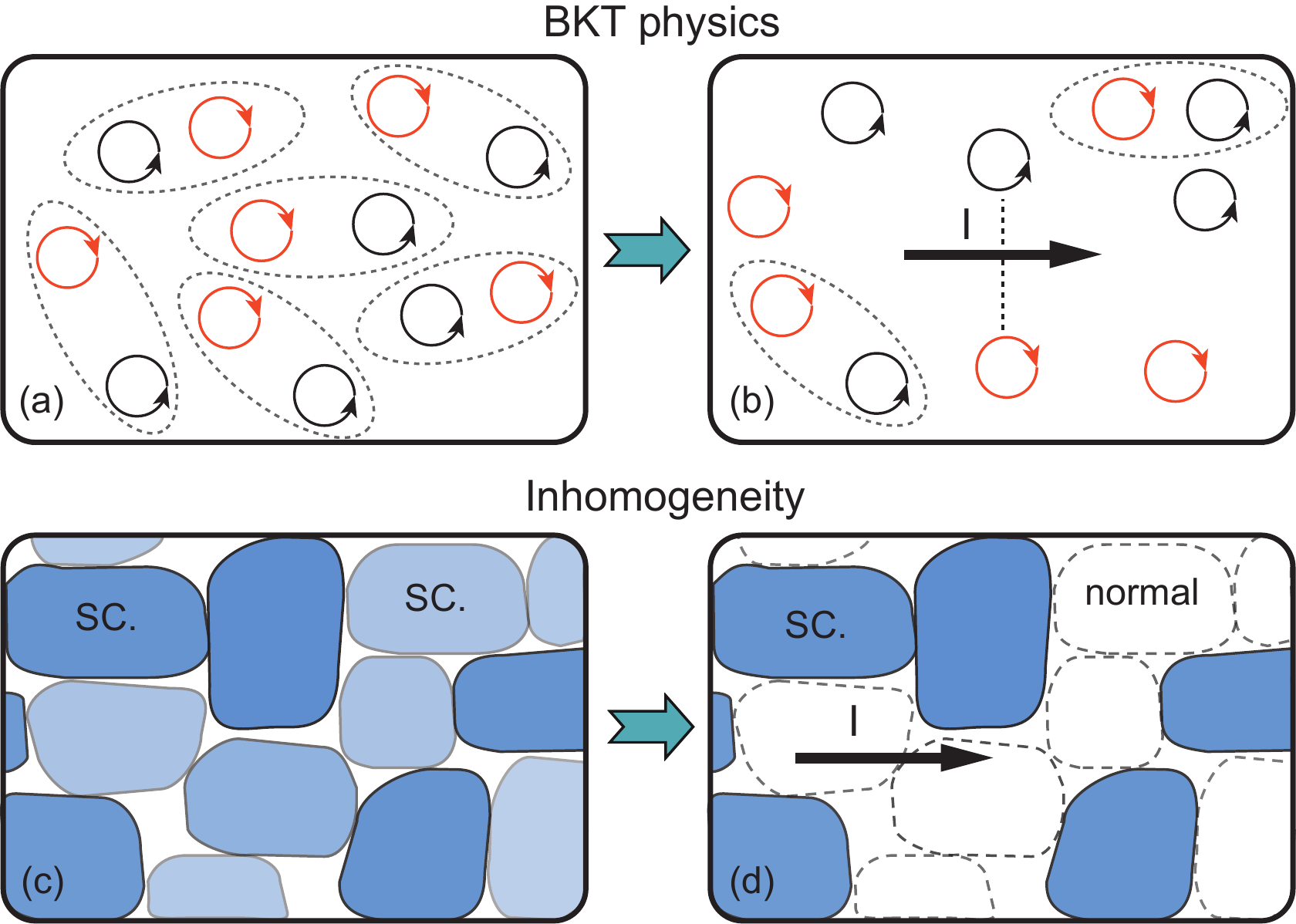}
\caption{Generation of non-linear $IV$ characteristics due to BKT physics (upper panels) and inhomogeneity (lower panels). 
In the BKT case the vortices, which are bound below $T_c$ in pairs with opposite vorticity (panel a), get unbound by 
a sufficiently large current $I$ (panel b). This generates an extra voltage drop proportional to the average density 
of unbound vortices, leading to non-linear characteristics. In the case of inhomogeneous superconductors instead the 
system segregates into puddles with different strength of the local SC condensate (panel c). As a consequence, a 
finite applied current $I$ can turn weak SC puddles into normal ones (panel d), non-linearly increasing the 
global resistivity.}
\label{fig:sketch}
\end{figure}

In this paper we analyze the role of SC inhomogeneity in the non-linear $IV$ characteristics of 
2D superconductors. We compare two paradigmatic systems: NbN thin films and STO-based interfaces. In the former 
case we show that the superfluid-stiffness behavior extracted from the measurements of the $IV$ characteristics 
is consistent with the direct measurements of $\lambda^{-2}$, and both are compatible with a BKT 
transition, even if the BKT universal jump is smeared by disorder 
\cite{mondal_bkt_prl11,yong_prb13,benfatto_review14,maccari_prb17}. In contrast, for 
STO-based interfaces the non-linearity of the $IV$ characteristics cannot be simply ascribed to vortex-antivortex 
unbinding triggered by a large current, as it happens within the BKT scheme, since this would lead to dramatically 
overestimate the BKT transition temperature. We then argue that in these systems the non-linearity of the $IV$ 
characteristics is due to the pair-breaking effect in the weaker SC regions, as the driving current increases, see 
Fig.\,\ref{fig:sketch}. By modelling this mechanism within an effective medium (EM) theory, we can reproduce a $IV$ 
non-linearity in qualitative agreement with the experiments, suggesting that mesoscopic inhomogeneity can essentially 
hinder the observation of BKT effects at these interfaces.

The plan of the paper is the following. In Sec.\ \ref{exp} we show the experimental results for the $IV$ characteristics in two paradigmatic cases, a NbN thin film and a STO-based sample. While in the former case a paradigmatic example of BKT physcis is found, in the latter pursuing a BKT analysis lead to clear inconsistencies. In Sec.\ \ref{theo} we then discuss an alternative model to explain the observed non-linearity in STO-based systems, and we compare it with the experiments. Sec. \ref{theo} contains the concluding remarks. In Appendix A we give more details on the penetration-depth measurements in the NbN sample, and Appendix B contains additional information on the theoretical model of Sec.\ \ref{theo}.

\section{Experiments}
\label{exp}

 \begin{figure*}[t]
\hspace{-1cm}
\includegraphics[width=0.7\textwidth,angle=-90]{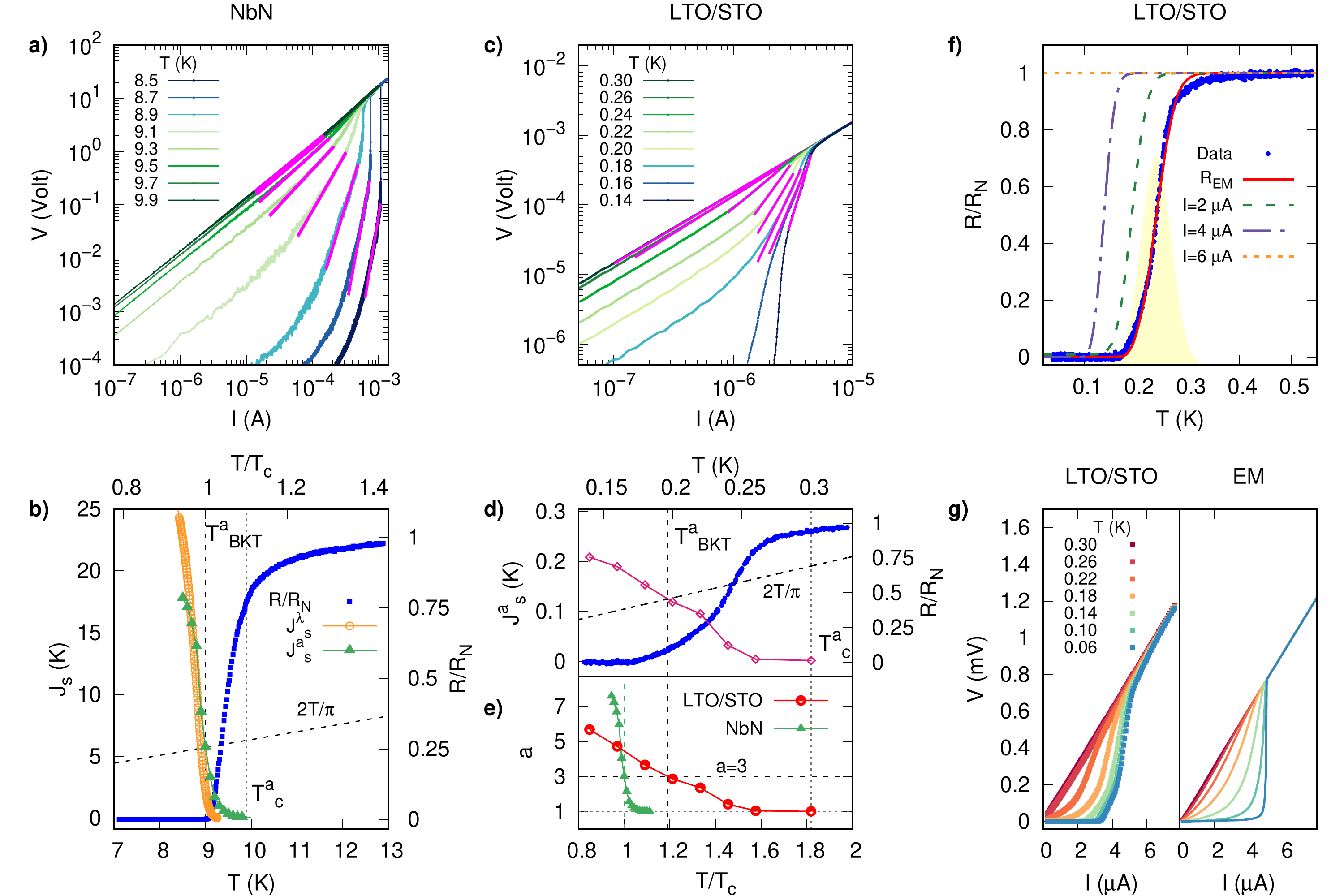}
\caption{(a)-(b) Experimental results for a 3\,nm thick NbN film. (a) Measurements of the $IV$ characteristics. 
Solid lines are fit with the Eq.\, 
\pref{abkt}. (b) Temperature dependence of the measured superfluid stiffness $J_s^\lambda$, compared with the one 
extracted from $IV$ characteristics, $J_s^a$, and with the normalized resistivity (right axis). The universal  
$2T/\pi$ BKT line is also shown. The slightly larger value of the SC transition $T_c^a$ extracted 
from $J_s^a$ can be ascribed to finite-size effects. (c)-(e) Experimental results for a LTO/STO sample. 
(c) Measurements of the $IV$ characteristics. Solid lines are fit with the Eq.\, \pref{abkt}. 
The resulting exponent $a(T)$ is shown in panel (e), and the corresponding stiffness $J_s^a$ in panel (d), 
along with the normalized sheet resistance. In panel (e) we show for comparison also $a(T)$ for NbN 
as a function of the reduced temperature $T/T_c$. (f)-(g) Comparison between the LTO/STO data and the 
theoretical results obtained within the EM approximation. (f) Normalised sheet resistance $R/R_N$ (blue dots) 
compared with the EM resistivity $R_{em}$ obtained from the numerical solution of Eq.\ \pref{emt} 
at $I=0$ (solid red line) and at finite $I$ (dashed lines). In background we show the probability 
distribution of  $T_c^i$, with $\bar T_c=0.24$\,K and $\sigma=0.029$\,K. (g) Experimental (left) 
and theoretical (right) $V(I)$ curves at different temperatures. For the EM calculations we used 
a larger variance $\sigma=0.06$\,K and $I_{c,0}=5$\,$\mu$A.}
\label{fig2}
\end{figure*}

Let us start with the case of a 3 nm thick NbN film grown on single crystalline MgO substrate. Details of sample preparation are given in \cite{mondal_bkt_prl11}. The $IV$ measurements were performed by means of a standard 4-probe technique, by using a current source and a nanovoltmeter in a conventional $^4$He cryostat where the sample is kept in contact with helium exchange gas to minimize heating effects. The temperature variation in all $IV$ scans was less than 30 mK. To improve sensitivity, the film was patterned into a 20 $\m$m wide stripline using ion-beam milling with large current contacts and narrow voltage contacts.
 The $IV$ characteristics at selected temperatures are shown in Fig.\,\ref{fig2}a, while in Fig.\ \ref{supp1}a of Appendix A we report the full data set. 
 
As mentioned above, within the BKT scenario the $IV$ characteristics acquire a non-linear dependence near $T_c$, since 
a large enough current can unbind the vortex-antivortex pairs present below $T_c$. This effect generates a voltage 
$V\propto n_V(I)I$, where the equilibrium density of free vortices $n_V(I)$ scales with a power-law of the applied current, 
with an exponent proportional to $J_s$ \cite{hn79}:
\begin{equation}
\label{abkt}
V\propto I^{a(T)}, \quad a(T)=1+\frac{\pi J_s(T)}{T}.
\end{equation}
In the ideal BKT case \cite{bkt,bkt2,nelson}  $J_s(T)$ is expected to jump discontinously at the intersection with the BKT line:
\be
\label{jump}
J_s(T^-_{BKT})=\frac{2}{\pi}T_{BKT}, \quad J_s(T^+_{BKT})=0.
\ee
When inserted into Eq.\ \pref{abkt}, this implies that also  the $IV$ exponent should jump at the transition:
\be
a(T_{BKT}^-)=3, \quad a(T^+_{BKT})=1.
\ee
In real 2D superconductors, as NbN thin films, the 
$J_s^\lambda$ obtained by direct measurements of $\lambda^{-2}$ by means of two-coil mutual inductance 
technique \cite{kamlapure_apl10,mondal_bkt_prl11,pratap_prb17} goes continuolsy to zero, but around the intersection with the universal BKT line it shows a rapid downturn with respect to the BCS temperature dependence,   see Fig.\ \ref{fig2}b and Fig.\ \ref{supp1}b of Appendix A. As discussed in previous work\cite{mondal_bkt_prl11,yong_prb13,benfatto_review14}, the low vortex-core energy and a moderate inhomogeneity of the sample account rather well for the smearing of the BKT jump. However, in this situation the  BKT temperature cannot be identified by the intersection of $J_s(T)$ with the universal BKT line, but it is defined by the real $T_c$ where $J_s^\lambda=0$, and $R(T_c)=0$. Analogously, if we denote by $J_s^a$ the stiffness extracted from the $IV$ characteristics, the critical temperature $T_c^a$ corresponds to the scale where $J_s^a(T_c^a)=0$ so that $a(T_c^a)=1$:
\be
\label{tca}
J_s^a(T)\equiv \frac{(a-1)T}{\pi},  \quad J_s^a(T_c^a)=0 \Ra a(T_c^a)=1.
\ee
 In Fig.\,\ref{fig2}b we show that in NbN $J_s^a$ closely matches $J_s^\lambda(T)$ below $T_c$, and vanishes at a slightly larger temperature.  This phenomenon can be ascribed to finite-size effects, since the current used to estimate $J_s^a$ from 
Eq.\,\pref{abkt} sets a finite length scale which rounds off the vanishing of the stiffness above $T_{BKT}$ 
\cite{bkt1}. This is the same effect usually seen while measuring the stiffness at finite microwave frequencies 
\cite{armitage_prb07,armitage_prb11,ganguly_prb15}. Thus, the critical temperature $T_c^a$ turns out to be few percent larger then the true $T_c$ set by dc transport, $R(T_c)=0$, or by 
the vanishing of $J_s^\lambda$. We also notice that the temperature $T_{BKT}^a$ where $a(T_{BKT}^a)=3$ has no 
particular significance in the realistic case of a smeared jump, but it is still expected to be lower than the real 
$T_c$. Finally, a word of caution concerns possible screening effects due to supercurrents. A crucial prerequisite for the occurrence of the BKT transition is that vortexes interact logarithmically at all length scales\cite{bkt,bkt1,bkt2}.
However, while this always happens in neutral superfluids, in charged superconductors the screening currents around the vortex core screen out the inter-vortices interactions at a scale sets (in 2D) by the Pearl length\cite{pearl} $\Lambda=2\lambda^2/d$. In order to see BKT physics one should then verify that $\Lambda$ is of the order of the sample size when the $J_s^\lambda$ downturn occurs. As shown explicitly in Appendix A, at the intersection with the 
universal BKT line   $\Lambda\approx 2.6$\,mm  is of the order of the size of our NbN sample, so screening effects are irrelevant.  On general ground, this condition usually occurs in thin enough films\cite{beasly_prl79}  since both the decrease of $d$ and the increase of $\lambda$ due to effectively higher disorder contribute to enhance the Pearl length. 

We now turn to the case of STO-based interfaces. Here we used a 10 u.c thick LaTiO$_3$ epitaxial layer grown on a TiO$_2$-terminated SrTiO$_3$ single by Pulsed Laser Deposition\cite{biscaras_natcomm10}.  The 3$\times$3 mm LTO/STO sample was thermally anchored to the last stage of a dilution refrigerator and standard four probes resistivity measurements were performed in a Van der Pauw geometry. Fig.\,\ref{fig2}c shows the $IV$ characteristics of a LTO/STO sample. The first observation is the presence of 
a persistent non-linear behavior over a wide temperature range {\em above} $T_c$, which is identified by the vanishing 
of the dc resistivity. 
This has to be contrasted with the case of NbN, where at $T\simeq 1.1\,T_c$ the $IV$  characteristics display a full linear behavior, as indeed expected in the metallic case where vortices are already thermally unbound.  
By 
closer inspection of Fig.\,\ref{fig2}c one sees also that the non-linear regime connects smoothly to 
the linear one, while for NbN in Fig.\,\ref{fig2}a the non-linear regime is followed by an abrupt jump at the 
critical current where normal-state resistance is recovered. Such a difference is due to the fact that in LTO/STO one 
is in practice observing non-linear characteristics {\em above} $T_c$, where no SC critical current exists but the 
resistivity is strongly temperature dependent. These features of the $IV$ characteristics and the 
consequent persistence of $J_s^a(T)$ above $T_c$ are very common in the literature in several reports for 
STO-based interfaces \cite{triscone_science07,han_apl14,caviglia_prb17} and other gated 2D superconductors 
\cite{ye_science15,castroneto_nat16,pasupathy_natphys16,herrera_nature18}, in particular for samples which show a 
considerable broadening of the resistive transition. 

Even though these observations should already suggest that at different mechanism is at play here, 
we can nonetheless pursue the BKT analysis based on Eq.\,\pref{abkt}, and extract the $a(T)$ exponent, see 
Fig.\,\ref{fig2}d. Even though some uncertainty in the determination of $a(T)$ stems from the 
limited fitting range, a robust finding is the persistence of the corresponding stiffness $J_s^a(T)$ far above $T_c$, see Fig.\,\ref{fig2}d. In particular, the critical temperature $T_c^a$ estimated from the $a$ exponent, see Eq.\ \pref{tca}, is almost {\em twice} as large as $T_c$, and even the temperature 
$T^a_{BKT}$ where $J^a_s$ intersects the BKT line is well above $T_c$. As discussed before in the case of NbN, a moderate shift of  $T_c^a$ with respect to  $T_c$ can be expected within a BKT scenario, as due to the fact that one probes the stiffness at the finite length scale set by the large current. However, while this can explain a ten percent increase of the critical temperature extracted from the $a(T)$ exponent in NbN, $T_c^a\simeq 1.1\,T_c$, it cannot account for $T_c^a\simeq 1.8\ T_c$ estimated from a BKT analysis of the LTO/STO sample. We notice that similar results have been found in previous attempts to interpret non-linear $IV$ characteristics in STO-based interfaces and other gated 2D superconductors\cite{triscone_science07,han_apl14,caviglia_prb17, ye_science15,castroneto_nat16,pasupathy_natphys16,herrera_nature18}, suggesting a common origin for the emergent non-linearity in confined systems.



\section{Role of Inhomogeneity}
\label{theo}
To explain the $IV$ non-linearity in LTO/STO we then propose a simple model, starting from the basic idea that in 
these systems transport is dominated by percolation through a strongly inhomogeneous background emerging at mesoscopic 
length scales \cite{biscaras_natmat13,bid_prb16,jespersen_prb16,scopigno2016,caviglia_natcomm18,hurand_prb19}. 
As already observed before 
\cite{benfatto_prb09,caprara_prb11,caprara_sust15}, a first signature of this inhomogenity is the observation of a marked broadening of the resistive transition. This finding cannot be ascribed 
to usual paraconductivity effects due to SC fluctuations, but it can be well captured by assuming that the 
metal-to-superconductor transition can be mapped onto a random-resistor-network problem. More specifically, we consider a set 
of local resistances $R_i$ which switch off from the normal-state value $R_N$ to zero at a local temperature 
$T_c^i$, whenever the driving current $I$ is below a threshold $I_c^i$. The local $T_c^i$ are distributed with a 
probability $P(T^i_c)$, with overall weight $w_s=\int dT_c^i P(T^i_c)$. The SC transition can be well understood 
already in the EM approximation \cite{caprara_prb11,caprara_sust15}, where the sample resistance 
$R_{em}(T,I)$ is a solution of the self-consistency equation \cite{landauer, kirkpatrick}
\begin{equation}
\lb{emt}
\sum_i \frac{R_i-R_{em}}{R_i+R_{em}}=0,
\end{equation}
where each $R_i$ has a probability  $w(T)=\int_T^\infty dT_c^i P(T_c^i)$ of being zero.
Even though the EM approach neglects spatial correlations, nonetheless it gives
insight about the qualitative behaviour of the system. At $I=0$ the condition $R_{em}=0$ requires that 
the fraction $w(T)$ of SC links has reached the percolation threshold $w^*$ (in two dimensions, $w^*=0.5$, 
see Appendix B for more details). The shape of $R_{em}(T,I=0)$ depends on the width of the $P(T_c^i)$ 
distribution, 
that sets the width of the paraconductivity regime, and on the total fraction $w_s$ of SC links. When $w_s$ is smaller 
than one, i.e., part of the system remains metallic, and slightly larger than the percolation threshold, i.e., 
$w_s\gtrsim 0.5$, one finds \cite{caprara_prb11} that $R_{em}$ has a marked tail above $T_c$, as shown by 
the numerical solution of Eq.\,\pref{emt} in Fig.\,\ref{fig2}f, in agreement with the experiments. Here, we assumed 
that the $T_c^i$ distribution is gaussian, with average $\bar T_c=0.24$\,K and standard deviation 
$\sigma=0.029$\,K, and we used $w_s=0.52$. As a consequence, when the temperature decreases below 
$\bar T_c+3\sigma\simeq 0.29$\,K the condition $T<T_c^i$ is fulfilled for a progressively larger fraction of 
local resistors $R_i$, which then switch off to zero, leading to a suppression of $R_{em}(T,I=0)$. 

A finite driving current is then able to break the weak links between the good SC regions having mesoscopic length 
scales.  Even though we lack a precise information on the nature of the microscopic weak links, we checked (see Appendix B) 
that the experimental data can be well reproduced by a temperature-dependent critical-current that follows the 
Ambegaokar-Baratoff formula \cite{amb-bar}:
\begin{equation}
\lb{ici}
I_c^i = I_{c,0} \frac{\Delta_i(T)}{\Delta_i(0)} \tanh{\left(\frac{\Delta_i(T)}{2 k_B T}\right)}
\end{equation}
where  $I_{c,0}$ at $T=0$ is independent of the resistor, and the $T=0$ value of the local gap scales 
with the local $T_c^i$ as ${\Delta_i(0)}/{k_B T_c^i} \approx 1.76$. The temperature dependence of 
$I_c$ following from Eq.\,\pref{ici} is also in good agreement with a recent analysis of the critical current in 
STO-based interfaces \cite{jespersen_prb16,hurand_prb19}, even though we cannot exclude a-priori that different models for SC weak links 
\cite{review_likharev} could work as well, as long as they reproduce the strong dependence of $I_c^i$ on the local 
$T_c^i$ at $T\simeq T_c^i$, see Appendix B.
From Eq.\,\pref{ici} we see that, at a given temperature, only the resistors having $I_c^i$ larger than the driving 
current $I$ can be SC, shifting the $R_{em}(T,I)$ curves towards lower temperatures, see Fig.\,\ref{fig2}f. The same 
effect is also responsible for the observed non-linearity of the $IV$ characteristics shown in Fig.\,\ref{fig2}g. 
Here, we used  $I_{c,0}=5$\,$\m$A and a 
somehow larger width $\sigma=0.06$\,K of the $P(T_c^i)$ distribution. Despite the simplifications implicit in our 
model, with this set of parameters we can very well reproduce the experimental curves. Below 
$T_c$ all the $I_c^i$ rapidly collapse towards $I_{c,0}$, which essentially identifies the real critical 
current in the SC state, see Appendix B. On the other hand, above $T_c$, in the whole regime of temperatures where 
$R_{em}(T,I=0)\ll R_N$ because of the sample inhomogeneity, the non-linear behavior is due to the current-induced 
breaking of the SC links. As $I$ increases a larger fraction of the SC links becomes normal,  and the global 
resistivity progressively crosses over towards its normal-state value. 
The wider distribution of $T^i_c$ found in the analysis of $IV$ characteristics may be ascribed to 
avalanche effects, not captured by our simple model. Indeed, after the 
first weak links break down more current flows in the remaining ones, which then will be easier 
to break and so on. As a consequence, the distribution of local SC links can get broader at larger applied currents, 
as indeed we found while comparing the theoretical simulation with the experiments.

\section{Conclusions}
In summary, we analyzed the $IV$ characteristics in two paradigmatic examples of 2D superconductors: NbN
thin films, and STO-based oxide interfaces. In the former case we observed a non-linear behavior  
well consistent with the typical occurrence of BKT physics in realistic systems. In particular, even if the universal BKT jump of the superfluid stiffness $J_s(T)$ is partly hindered by nanoscopic 
inhomogeneity of the SC background, its essential features remain visible and reflect in a similar fashion on  $J_s^\lambda$ extracted by measurements of the inverse penetration depth, or on $J_s^a$ extracted by measurements of $IV$ characteristics. In the presence of a smeared jump, the intersection of $J_s(T)$ with the universal $2T/\pi$ BKT line has no particular meaning, and the critical temperature is identified by the temperature scale where the superfluid stiffness effectively vanishes. For $J_s^\lambda$ this occurs exactly at the same $T_c$ where the resistivity goes to zero. In the case of $J_s^a$ we observed a few percent increase of $T_c^a$ with respect to $T_c$, that we ascribed to finite-size effects. Indeed, in full analogy with what observed measuring the stiffness at finite frequencies\cite{armitage_prb07,armitage_prb11,ganguly_prb15}, the stiffness probed at the reduced length scale set by the finite current appears finite in a small range of temperatures above the real $T_c$, set by the vanishing of the large-distance superfluid rigidity. When rephrased in term of the $IV$ critical exponent $a(T)$, this implies that one should not focus on the scale where $a=3$, that corresponds to the intersection with the universal $2T/\pi$ line, but with the scale $T_c^a$ where $a=1$, and compare it with the $T_c$ where global resistance vanishes. 

In the case of STO-based  interfaces we argued that the
non-linear $IV$ characteristics cannot be ascribed to a BKT phenomenon, but rather to the existence of a strong 
fragmentation of the SC properties on mesoscopic length scales. On the experimental side, we identified two typical signatures of the emergent inhomogeneity: a marked rounding of the resistance, that cannot be explained with usual paraconductvity effects\cite{benfatto_prb09,caprara_prb11,caprara_sust15}, and an estimate of $T_c^a$ extracted from a BKT-like fit of the $IV$ characteristics almost twice as large as the $T_c$ where $R=0$. This result implies that the non-linear behavior emerges mostly {\em above} the resistive transition temperature $T_c$. Even though the direct measurement of $J_s^\lambda$ in our sample is not available, due to the fact that it would require a dedicated microwave setup\cite{bergeal_natcomm18,caviglia_cm18}, these findings suggest an alternative origin for the observed non-linear transport. We then showed that by modelling the SC transition by means of a random-resistor network we can well reproduce both the rounding of the resistive transition and the emergence of non-linear characteristics. The basic idea is that transport occurs via a network of metallic and SC regions, whose fraction depends both on the temperature and on the driving current. By assuming a distribution of the local SC temperatures $T_c^i$ and critical current $I_c^i$ the global resistivity of the sample, computed by means of an effective-medium approximation, is progressively lowered as the temperature decreases towards $T_c$, where the SC fraction reaches the percolation threshold and the superfluid transition occurs. As the current increases, it can overcome the local critical current $I_c^i$, reducing the overall SC fraction and leading to an increase of the resistance, that manifests with non-linear $IV$ characteristics. 
As we discussed in the introduction, there have been several indirect evidences that the SC background in STO-based interfaces fragments in islands of about one hundred of nanometers \cite{biscaras_natmat13,bid_prb16,jespersen_prb16,scopigno2016,caviglia_natcomm18,hurand_prb19}. These could be due to intrinsic effects, like an electronic phase separation, as due either to the non-rigidity of the band structure at the 
interfacial potential well \cite{scopigno2016}, or to a strong density-dependent Rashba spin-orbit coupling 
\cite{caprara_prl12}. On the other hand, also extrinsic effects can play a role and cooperate in the formation of a widely fragmented SC landscape. For example, it has been recently observed that one-dimensional like superconductivity can be triggered by the domain structures in the STO substrate\cite{moler_natmat13,moler_prb16,kalisky_natmat17,kalisky_prb17}, leading to modulations on much larger length scales, of order of tens of micrometers. While we cannot exclude that these stripy features contribute to the observed non-linear transport, it is worth noting that an "apparent" BKT behavior discussed in the of the $IV$ characteristics similar to the one 
observed in our LTO/STO sample is very common in the literature, especially for gated superconductors. Indeed, 
it has been seen in other STO-based interfaces \cite{triscone_science07,han_apl14,caviglia_prb17}, in 
2D transition-metal dichalcogenides 
\cite{ye_science15,castroneto_nat16,pasupathy_natphys16,dezi} and also in the recently discovered twisted bilayer 
graphene \cite{herrera_nature18}. As a consequence, while our results question 
the possibility to observe a BKT physics in this extremely confined 2D electron gas, they also suggest that 
non-linear $IV$ characteristics can be used as a benchmark for emergent inhomogeneity in a wide class of 
superconductors.

\vspace{0.5cm}
{\em Acknowledgements} 

The work was supported  by Italia-India collaborative project SuperTop (Italian MAECI PGRO4879 and Indian 
Department of Science and Technology No. INT/Italy/P-21/2016 (SP)), by the Sapienza University via Ateneo 2017 
(prot.  RM11715C642E8370) and  Ateneo 2018 (prot.   RM11816431DBA5AF), by the Department of Atomic Energy, Govt. 
of India, Department of Science and Technology, Govt. of India (Grant No: EMR/2015/000083), by the IFCPAR 
French-Indian program (Contract No. 4704-A), by the Delegation G\'en'erale \`a l'Armement (which supported the 
PhD grant of SH), and by the Nano-SO2DEG project of the JCJC program of the ANR.

\vspace{0.5cm}
{\em Authors contribution}
MM, JJ and PR synthesized the NbN sample and performed the measurements on it, AD and RCB provided the LTO/STO sample and JB, SH, NB and JL performed the measurements on it, GV, SC and LB elaborated the theoretical model and 
GV performed the numerical calculations. LB conceived the project together with SC, NB and PR. LB wrote the manuscript  
with inputs from all the coauthors. 

\appendix
\section{Measurements of $1/\lambda^2$ in NbN}

\begin{figure*}[t]
\includegraphics[width=0.5\textwidth,angle=-90]{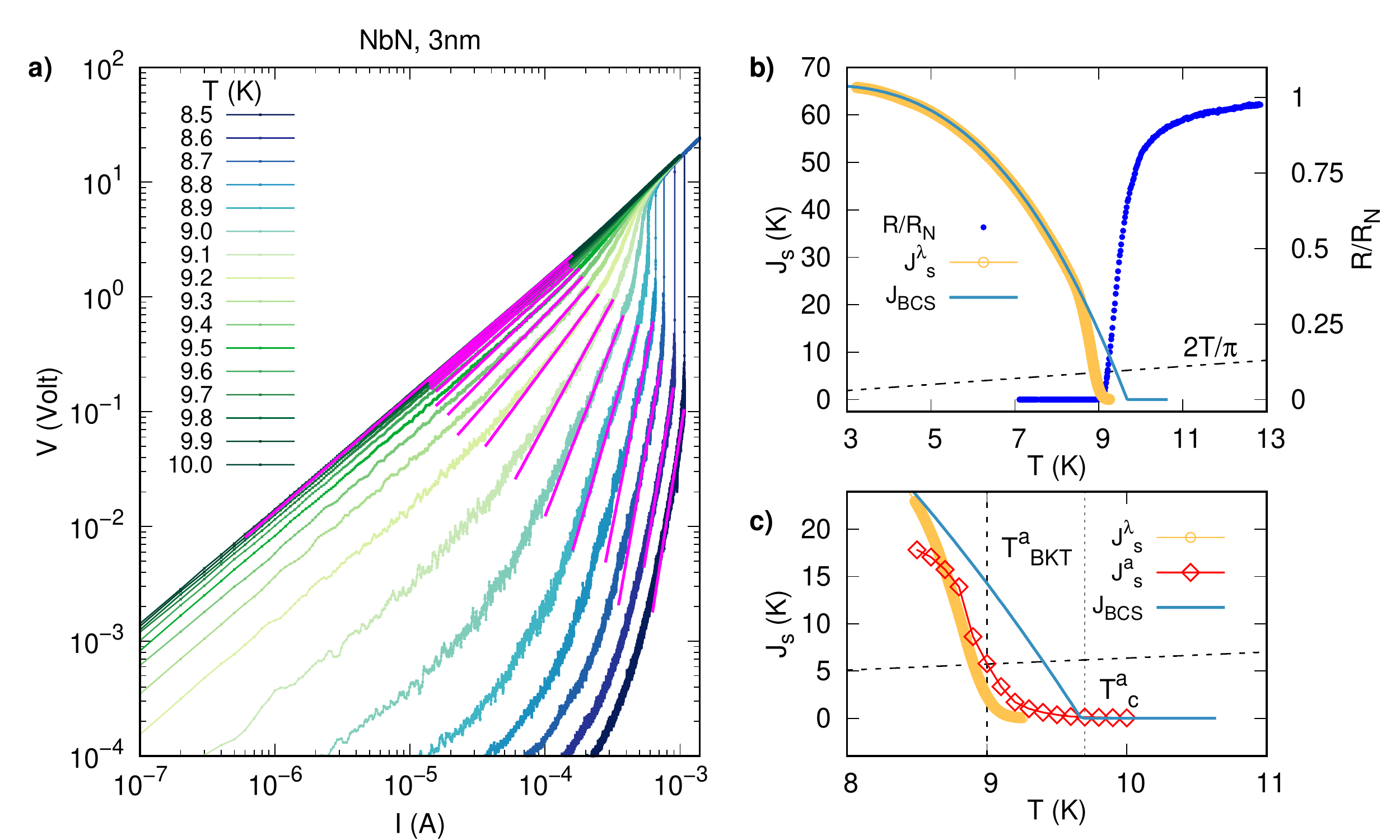}
\caption{ (a) Measurements of the $IV$ characteristics for our 3nm NbN thin film. The same data at selected temperatures are shown in Fig.\ \ref{fig2}a. Solid lines are fit with the Eq.\ \pref{abkt}. (b) Temperature dependence of the superfluid stiffness $J_s^\lambda$ obtained from the measurement of the penetration depth, along with its BCS fit $J_{BCS}$ based on Eq.\ \pref{bcs}. Here we used $\Delta(0)/T_{MF}=2.2$, that is slightly larger than the weak-coupling BCS limit, as already observed in the systematic analysis of superfluid stiffness in NbN\cite{mondal_bkt_prl11,yong_prb13}. As expected, the mean-field temperature is larger than the real critical temperature $T_c$, identified by the vanishing of the sample resistance (right axis). (c) Enlarged view near the transition, where we also show the $J_s^a$ corresponding to $a(T)$ exponent obtained from the fit of $IV$ curves in panel (a).}
\label{supp1}
\end{figure*}

In Fig.\ \ref{supp1}a we report the full set of $IV$ characteristics along with the fit based on Eq.\ \pref{abkt}. In Fig.\ \ref{supp1}b we show in an extended range the measurements of the inverse penetration depth in our NbN film. Details of the two-coils mutual inductance  measurements can be found in \cite{kamlapure_apl10,mondal_bkt_prl11,pratap_prb17}. The measured $1/\lambda^2$ has been converted in the stiffness energy scale $J_s^\lambda$ by means of the standard relation:
\be
J_s^\lambda[K]=6.2\times \frac{d[nm]}{\lambda^2[\m m^2]}.
\ee

As shown in Ref.\ \cite{mondal_bkt_prl11,yong_prb13}, $J_s^{\lambda}$ closely follows at low temperatures the BCS temperature dependence. This is explicitly shows in Fig.\ \ref{supp1}b, where we compare the superfluid stiffness $J_s^{\lambda}$ with the BCS fit $J_{BCS}$, based on the following expression:
\be
\lb{bcs}
\frac{J_{BCS}}{J_{BCS}(0)}=\frac{\Delta(T)}{\Delta(0)}\tanh (\Delta(T)/2T),
\ee
where $\Delta(T)/\Delta(0)$ is computed from the self-consistent BCS equation, and vanishes at the mean-field temperature $T_{MF}$. Both $T_{MF}$ and $\Delta(0)/T_{MF}$ are obtained by the fit of $J_s^\lambda$ at low temperature. As one can see in Fig.\ \ref{supp1}b, the BCS fit accurately reproduce the data up to  $T\simeq 8.5$ K, where a rapid downturn due due to vortex unbinding start to be visible. By accounting for a moderated inhomogeneity of the sample, and for the small vortex-core energy, one can indeed identify this downturn with the universal BKT jump, smeared by disorder\cite{mondal_bkt_prl11,yong_prb13}. 

Notice that in our film finite-size effects due to screening currents are irrelevant near $T_c$. 
At the intersection with the BKT line $J_s\simeq 5$ K so that $\lambda\simeq 2$ $\m$m. As a consequence the Pearl length\cite{pearl} $\Lambda=2\lambda^2/d\simeq $ 2.6 mm. Since our sample is around 8 mm in diameter, we are safely in the condition where screening effects due to charged supercurrent can be neglected. In addition, it is worth noting that screening effects act as a finite-size cutoff  for the logarithmically vanishing stiffness at the transition, so they would only give a smearing of the jump above $T_c$. What we observe is instead a rather symmetric smearing of the jump around the intersection with the universal line. As recently discussed in Ref.\ \cite{maccari_prb17}, this is a characteristic signature of the inhomogeneous SC background, which allows for vortex-pair proliferation already below $T_c$ in the bad SC regions.

\section{Theoretical model}
\subsection{The effective medium approximation for the random-resistor network}
As explained in Sec.\ \ref{theo}, to simulate the mesoscopic inhomogeneity in STO-based samples we describe the inhomogeneous SC background by means of a random resistor network (RRN) model. In this picture, every bond represents a resistor $R_i$, made by a mesoscopic region of electrons, with a specific local critical temperature $T^i_c$ randomly distributed. The global resistance $R_{em}$ of the system is given, within the effective-medium approximation (EMA), as a solution of Eq.\ \pref{emt}, where the sum is carried over all the bonds. An equivalent way to rewrite Eq.\ \pref{emt} is to sum instead over all possible values $\rho$ attained by the local resistors, weighted with the corresponding probability distribution $p(\rho)$:
\begin{equation} 
\label{eq:emt2}
\int p(\rho)  \frac{R_{em}-\rho}{R_{em}+ \rho}=0. 
\end{equation}
Suppose now that the each resistor can take only two constant values: $R_i=R_N$ if the link is in the normal-state, and $R_i=0$ if the temperature is lowered below the bond critical temperature $T_c^i$, so the temperature dependence in each bond will be $R_i=R_N \theta(T-T_c^i)$, where $\theta(x)$ is the Heavyside step function. If we denote with $P(T^i_c)$ the probability distribution of the local critical temperatures, the probability distribution of resistivity in Eq.\ \pref{eq:emt2} is then $p(\rho)=w(T)\delta(\rho)+[1-w(T)]\delta(\rho-R_N)$, where $w(T) \equiv \int_T^{+\infty}P(T^i_c)\,dT^i_c$ is the statistical weight of the superconducting fraction. Eq.\ \pref{eq:emt2} then reduces to:
\be
\lb{eq:emt3}
w+(1-w)\frac{R_{em}-R_N}{R_{em}+R_N}=0.
\ee
The critical temperature $T_c$ of the network, i.e. the temperature where $R_{em}\ra 0$,  is then  defined by Eq.\ \pref{eq:emt3} as the temperature where the SC fraction reaches the percolation threshold of $1/2$, as expected in two dimensions\cite{caprara_prb11}:
\begin{equation}
w(T_c)=\int_{T_c}^{+\infty}P(T_c)\,dT_c \equiv \frac{1}{2}.
\label{eq:Tc_alpha}
\end{equation}

For the distribution of local critical temperatures we assume a Gaussian distribution 
\begin{equation}
\label{eq:Tcdist}
P(T^i_c)=\frac{w_s}{\sqrt{2\pi}\sigma} e^{ -\frac{ (T^i_c-\overline{T}_c)^2 }{2\sigma^2} } 
\end{equation}
with average value $\overline{T}_c$ and variance $\s$, $w_s$ representing the total fraction of SC regions in the material. To determine numerically the EMA solution we will resort to the form \pref{emt}, by randomly sampling the local $T_c^i$ of each resistor according to the distribution \pref{eq:Tcdist}.  
At each temperature $T$ a fraction $\int_T^{\infty}dT_c^iP(T_c^i)$ of bonds are "switched-off", following the condition
\be
R_i= \begin{cases} 
1, & \mbox{if } T_c^i < T \\
0, & \mbox{if } T_c^i \geq T \\
\end{cases}
\ee
so that the effective resistivity $R_{em}$ will diminish by lowering the external temperature, until the percolation threshold $w=0.5$ is reached and $R_{em}$ becomes zeo. This procedure is more convenient than the numerical solution of Eq.\ \pref{eq:emt2} to implement the effects of a finite current, as we shall see in the next section.

\subsection{Effects of a finite current}
Starting from the EMA,  the information about the local critical temperature of each bond can be easily implemented as:
\begin{equation}
\label{eq:R(I)}
R_i(T, I)= \begin{cases} 
1, & \mbox{if }T \geq T_c^i, \\
0, & \mbox{if }T < T_c^i, I \leq I_c^i, \\
1, & \mbox{if }T < T_c^i, I > I_c ^i,  
\end{cases}
\end{equation}
where $I_c^i$ is the critical temperature of the $i$-th bond. 
In the absence of a full microscopic model for the SC puddles, we analyzed different critical-current schemes for the relation $I^i_c=f(T^i_c,T)$ and compared them with the data, in order to get an insight on the physical mechanism at play. The simplest relation one can guess is the Ginzburg-Landau (GL) relation for the critical current:
\begin{equation}
\label{eq:I_GL}
I^i_c=I^i_0(T) (T^i_c-T)^{3/2}.
\end{equation}
Here $I^i_0(T)$ sets the magnitude of the current, depending on the microscopic structure of the material; in principle, it can be a function of the external temperature $T$ and it can depend on the single resistor. As a starting point, we consider the easiest case $I^i_0(T)=I_0$ so the function is analytically invertible and therefore, for the $i$-th resistor to be superconducting, the condition to be fulfilled is $T_c^i \geq T + (I/I_0)^{2/3}$. We thus have 
\begin{equation}
R_i= \begin{cases} 
1, & \mbox{if } T_c^i < T_{eff}, \\
0, & \mbox{if } T_c^i \geq T_{eff}, \\
\end{cases}
\label{eq:R_cases_Teff}
\end{equation}
where $T_{eff}=T + \left(\frac{I}{I_0}\right)^{2/3}$ is the effective temperature perceived by the resistors. In this situation the $R_{em}$ depends on the applied current and the $IV$ characteristics will be in general non-linear.

\begin{figure}
\centering
\includegraphics[width=0.45\textwidth,angle=-90]{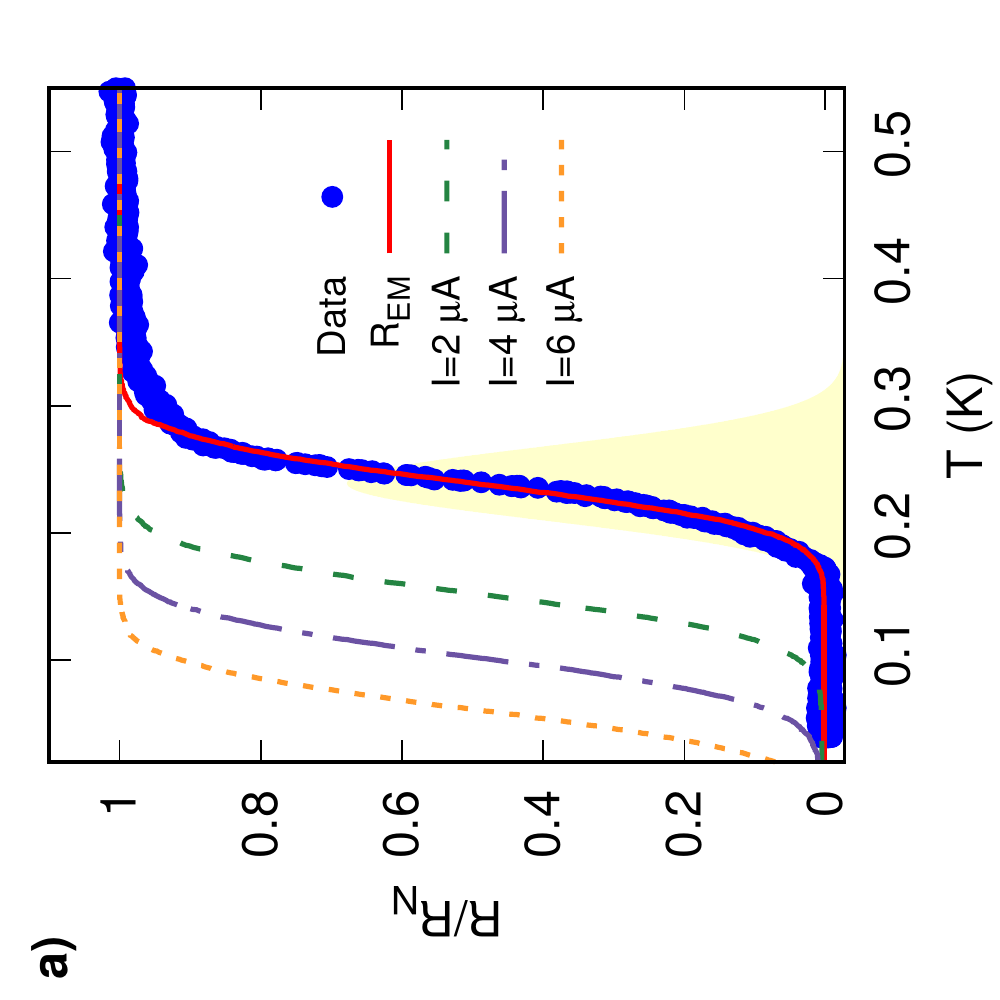}
\qquad
\includegraphics[width=0.45\textwidth,angle=-90]{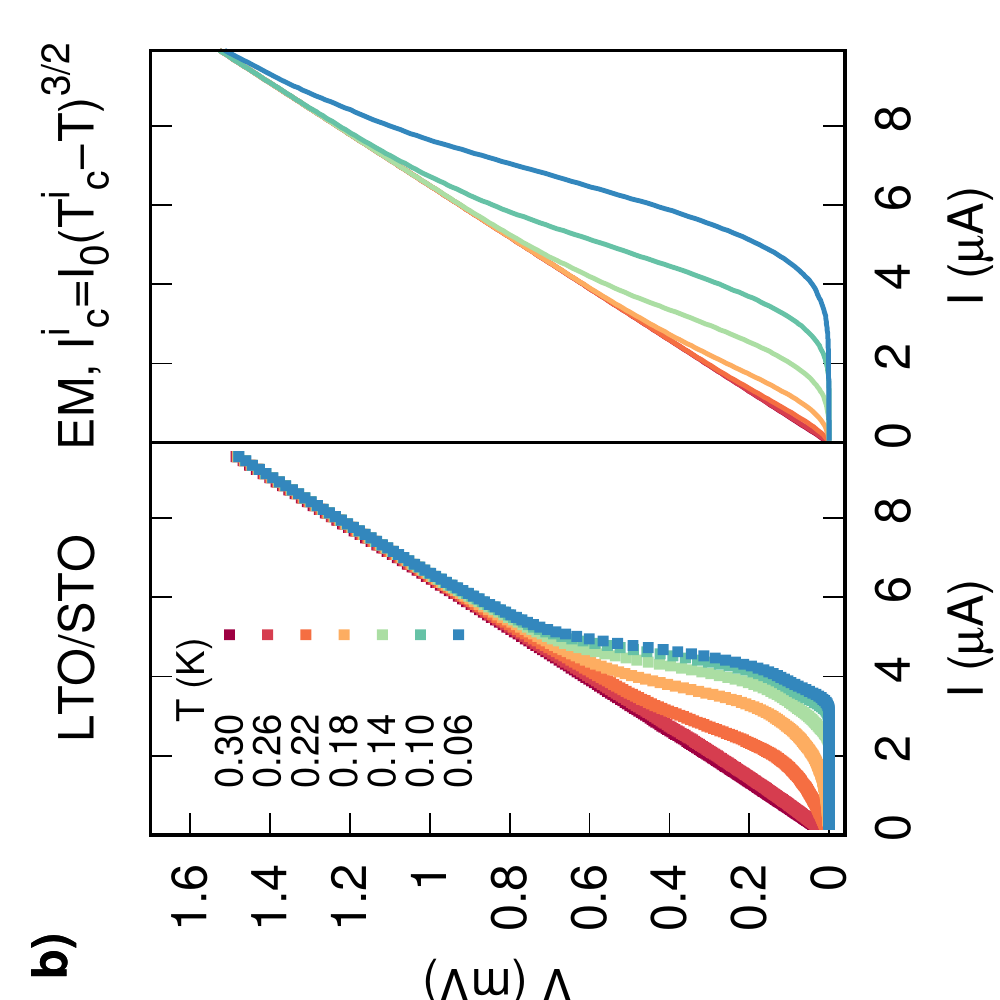}
\caption{(a) Normalised sheet resistance $R/R_N$ (blue dots) compared with the EM resistivity $R_{em}$ obtained from the numerical solution of Eq.\ \pref{emt}  at $I=0$ (solid red line) and at finite $I$ (dashed lines). In background we show the probability distribution of  $T_c^i$, with $w=0.5$, $\bar T_c=0.24$\,K and $\sigma=0.029$\,K. (b) Experimental (left) and theoretical (right) $V(I)$ curves at different temperatures using the GL relation\ \pref{eq:I_GL}.}
\label{R_IV_LG}
\end{figure}

In Fig.\ref{R_IV_LG} we show the resistivity curve and the $IV$ characteristics at different $T$ in the GL case.
The effective resistivity $R_{em}$ (solid red curve in fig. \ref{R_IV_LG}a) fits well the experimental data at vanishing driving current, using parameters $w=0.5$, $\s=0.029$ K, $\overline{T}_c=0.24$ K. At finite current, using { $I_0=80$ $\m$A}, we obtain the $R_{em}(T,I)$ displayed in Fig.\ \ref{R_IV_LG}a with dashed lines. Despite the fact that one obtains in general an increasing of $R_{em}$ as $I$ increases for a fixed temperature, the agreement with the experimental $IV$ curves is very poor. In Fig. \ref{R_IV_LG}b we compare the experimental $IV$ characteristics of our LTO/STO sample with the EMA numerical calculations.
The experimental data display a tendency to recover the ideal behaviour of a homogeneous superconductor as the temperature decreases, i.e. $V\propto I \th(I-I_c)$ when $T\ra 0^+$. This trend is not captured by the numerical calculation presented in the right panel of Fig. \ref{R_IV_LG}b, that provides very broad $IV$ characteristics, even at temperatures much lower than the percolation temperature $T_{perc}\simeq 0.19$K. To understand the origin of such drawback, we computed the probability distribution $P_I(I_c)$ of the critical currents, that is directly related to 
$P(T^i_c)$ by  $P_I(I_c)=\int \delta(I_c-f(T^i_c)) P(T^i_c) dT^i_c$, where  $I_c=f(T_c)$ is the functional relation between the local critical current and the local critical temperature. Given its inverse function  $T_c=g(I_c)$ one simply gets
\begin{equation}
\label{eq:P_I(I_c)}
P_I(I_c)=\frac{P(g(I_c))}{|f'(g(I_c))|}.
\end{equation}
where $P(x)$ is the distribution given in Eq. \eqref{eq:Tcdist}.

For the GL model of the critical current we showed above that $f(T_c)=I_0(T_c-T)^{3/2}$ and $g(I_c)=T_{eff}$, so that $P_I(I_c)$ takes the following form:
\be
\label{eq:P(I)_LG}
P_I(I_c)=\frac{2 w}{3 \s \sqrt{2\pi} I_0^{2/3}} \frac{e^{ -\frac{(I_c/I_0)^{4/3}}{2\s^2}}}{I_c^{1/3}}.
\ee
The main result is that in this case $P_I(I_c)$ does {\em not} depend on the external temperature $T$. This is also evident looking at the 
resistivity at finite $I$ in Fig. \ref{R_IV_LG}a, where all curve are obtained by shifting of the resistivity at $I=0$. This is a consequence of the fact that in the GL case the effect of the finite current is just to redefine the effective temperature of the system, as given by  Eq.\,\eqref{eq:R_cases_Teff}. In contrast, the experimental data 
shown in the left panel of Fig. \ref{R_IV_LG}b suggest that while above $T_c$ the system recovers smoothly the normal-state resistivity as $I$ increases, i.e. a wide distribution of local $I_c^i$ is present, as $T$ decreases the $V$ jumps almost suddenly to the normal-state value as $I$ increasing, signalling that the distribution of local $I_c$ values should progressively shrink towards a critical value $I_{c,0}$ that is the same for all the mesoscopic resistors.

\begin{figure*}
\centering
\includegraphics[width=0.5\textwidth,angle=-90]{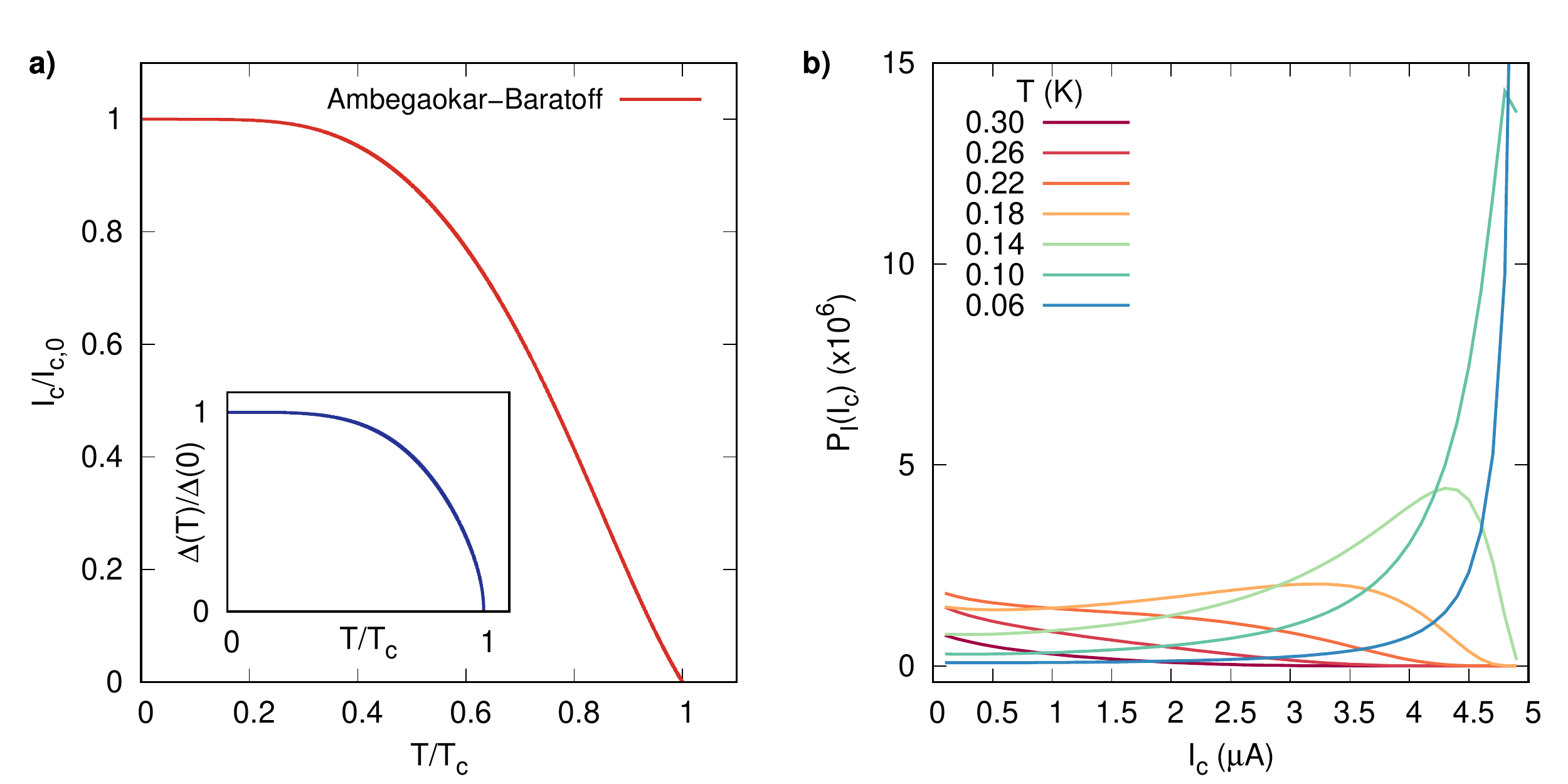}
\caption{a)  Temperature dependence of the critical current according to the Ambegaokar-Baratoff model \pref{eq:joseph}. Inset: approximated expression for the BCS-like temperature dependence of the gap, as given by Eq.\  \pref{eq:delta}.  (b) Probability distribution of the critical current for Ambegaokar-Baratoff model \pref{eq:joseph}, computed from Eq. \eqref{eq:P_I(I_c)}. The critical temperatures are distributed with the non-normalised Gaussian in Eq. \eqref{eq:Tcdist}, using the fitting parameters $w=0.52$, $\overline{T}_c=0.24$\,K, $\s=0.06$\,K and { $I_{c,0}=5$\,$\m$A}}
\label{fig:Amb_Bar}
\end{figure*}

These observations suggest that a different modelling for $I_c^i(T)$, able to satisfy two requirements:  (i) the zero temperature critical current must be independent on the single resistor $I^i_{c,0}=const$, (ii) the critical current should saturate pretty fast to its zero-temperature value in order to recover the behaviour of $IV$ curves at low temperature. The second item is also suggested by recent measurements in an other STO-based sample of the critical-current distribution below $T_c$\cite{hurand_prb19} . We then explored the outcomes of the Ambegaokar and Baratoff \cite{amb-bar} formulas, describing the critical current for a weak link between two SC electrodes
\begin{equation}
\label{eq:Amb-Bar}
I_c R_N = \frac{\pi\Delta(T)}{2e} \tanh\left(\frac{\Delta(T)}{2 k_B T}\right).
\end{equation}
%
According to Eq.\ \pref{eq:Amb-Bar} the critical current through a constriction scales with the superfluid density, that is expected to follow the BCS-like relation for $J_S$ reported in Eq.\ \pref{bcs} above. with 
$J_S(T)=J_S(0)\frac{\Delta(T)}{\Delta (0)}\tanh\left( \frac{\Delta (T)}{2 k_B T} \right)$. To mimic the BCS temperature dependence of the gap  $\Delta_i(T)$ in each resistor we use a simple approximated formula that reproduces well the BCS behavior (see inset of Fig. \ref{fig:Amb_Bar}a):
\begin{equation}
\lb{eq:delta}
f(\tau^i)=\frac{\Delta_i(T)}{\Delta_i (0)}=\left(1-\frac{\tau^4}{3}\right)\sqrt{1-\tau^4},\qquad
\frac{\Delta_i (0)}{k_B T^i_c} \simeq 1.76
\end{equation}
where $\tau^i=T/T^i_c$. The resulting temperature dependence of $I_c(T)$ from Eq.\ \pref{eq:Amb-Bar} is shown in Fig. \ref{fig:Amb_Bar}a.  As mentioned above, the experimental data suggest that all resistors have the same critical current as $T\ra 0$. We then assume for each local resistor the following temperature-dependent critical current:
\begin{equation}
\label{eq:joseph}
I_c^i(T)=I_{c,0}f(\tau^i)\tanh\left(\frac{1.76}{2}\frac{f(\tau^i)}{\tau^i}\right),
\end{equation}
corresponding to Eq.\ \pref{ici} above. In this case all the local link have the same $I_{c,0}$ as $T\ra 0$, but their behavior is different as $T$ approaches the local transition temperature $T_c^i$. The $IV$ characteristics obtained from the model \pref{eq:joseph} are shown in Fig. 2g. As one can see, they reproduce very well the experimental findings. In particular, the model \pref{eq:joseph} accounts for the sharpening of the RRN critical current as $T$ is lowered below $T_c$, as one can see in Fig.\  \ref{fig:Amb_Bar}b where we show the $P_I(I_c)$ obtained by inverting numerically the $T_c^i$ vs $I_c^i$ relation from Eq.\ \pref{eq:joseph}. Here one recovers a narrowing of the critical-current distribution as $T$ is lowered below $T_c\simeq 0.19$ K, and already for $T\simeq 0.06$ K $P_I(I_c)$ tends to a delta function centered at $I_{c,0}$.

\begin{thebibliography}{9}

\bibitem{reviewTMD}
Yu Saito, Tsutomu Nojima and Yoshihiro Iwasa, Nature Rev. Mat. {\bf 2}, 16094
(2016).

\bibitem{reviewSTO}
S. Gariglio, M. Gabay and J. M. Triscone, APL Materials {\bf 4}, 060701 (2016).

\bibitem{herrera_nature18}
Yuan Cao, Valla Fatemi, Shiang Fang, Kenji Watanabe, Takashi Taniguchi, Efthimios Kaxiras and Pablo Jarillo-Herrero, Nature {\bf  556}, 43 (2018).


\bibitem{bkt}
V.\ L.\ Berezinskii, Sov. Phys. JETP {\bf 34}, 610 (1972).

\bibitem{bkt1} J.\ M.\ Kosterlitz and D.\ J.\ Thouless, J. Phys. C {\bf 6}, 1181 (1973).

\bibitem{bkt2} J.\ M.\ Kosterlitz, J. Phys. C {\bf 7}, 1046 (1974).


\bibitem{beasly_prl79}
M.R. Beasley, J.E. Mooij and T.P. Orlando, \prl\, {\bf 2}, 1165 (1979).

\bibitem{nelson}
D.R. Nelson and J.M. Kosterlitz, Phys. Rev. Lett. {\bf 39},
1201 (1977).
%
\bibitem{hn79}
B. I. Halperin and D. R. Nelson, J. Low. Temp. Phys. {\bf 36}, 599 (1979).


\bibitem{epstein_prl81}
K. Epstein, A. M. Goldman, and A. M. Kadin, \prl \, {\bf 47}, 534 (1981).

\bibitem{epstein_prb83}
K. Epstein, A. M. Goldman, and A. M. Kadin, \prb \, {\bf 27}, 6691 (1983).

\bibitem{fiory_prb83}
A. T. Fiory, A. F. Hebard, and W. I. Glaberson, Phys. Rev. B {\bf28}, 5075 (1983).

\bibitem{lemberger_prl00}
S. J. Turneaure, T. R. Lemberger, and J. M. Graybeal, Phys. Rev. Lett. {\bf84}, 987 (2000).

\bibitem{armitage_prb07}
R.W. Crane, N. P. Armitage, A. Johansson, G. Sambandamurthy, D. Shahar, and G. Gruner, \prb~{\bf 75}, 094506 (2007).

\bibitem{armitage_prb11}
W. Liu, M. Kim, G. Sambandamurthy, and N.P. Armitage, 
Phys. Rev. B {\bf 84}, 024511 (2011).

\bibitem{kamlapure_apl10}
A. Kamlapure, M. Mondal, M. Chand, A. Mishra, 
J. Jesudasan, V. Bagwe, L. Benfatto, V. Tripathi, and
P. Raychaudhuri, Appl. Phys. Lett. {\bf 96}, 072509 (2010).

\bibitem{mondal_bkt_prl11}
M. Mondal, S. Kumar, M. Chand, A. Kamlapure, G. Saraswat, G. Seibold, L. Benfatto, and P. Raychaudhuri, 
Phys. Rev. Lett. {\bf 107}, 217003 (2011).


\bibitem{yazdani_prl13}
S. Misra, L. Urban, M. Kim, G. Sambandamurthy, and A. Yazdani, \prl~{\bf 110}, 037002 (2013).

\bibitem{yong_prb13}
Jie Yong, T. Lemberger, L. Benfatto, K. Ilin, M. Siegel,
Phys. Rev. B {\bf 87}, 184505 (2013).

\bibitem{ganguly_prb15}
Rini Ganguly, Dipanjan Chaudhuri, Pratap Raychaudhuri, Lara Benfatto,
Phys. Rev. B {\bf 91}, 054514 (2015).

\bibitem{bert_prb12}
Julie A. Bert, Katja C. Nowack, Beena Kalisky, Hilary Noad, John R. Kirtley, Chris Bell, Hiroki K. Sato, Masayuki Hosoda, Yasayuki Hikita, Harold Y. Hwang, and Kathryn A. Moler,
Phys. Rev. B {\bf 86}, 060503(R)  (2012).

\bibitem{bergeal_natcomm18}
G. Singh, A. Jouan, L. Benfatto, F. Couedo, P. Kumar, A. Dogra, R. Budhani, S. Caprara, M. Grilli, E. Lesne, A. Barthelemy, M. Bibes, C. Feuillet-Palma, J. Lesueur, N. Bergeal, Nat. Comm. {\bf 9}, 407 (2018).

\bibitem{caviglia_cm18}
 Nicola Manca, Daniel Bothner, Ana M. R. V. L. Monteiro, Dejan Davidovikj, Yildiz G. Saglam, Mark Jenkins, Marc Gabay, Gary Steele, and  Andrea D. Caviglia, Phys. Rev. Lett. {\bf 122}, 036801 (2019). 
 


\bibitem{coura_prb05}
G. M. Wysin, A. R. Pereira, I. A. Marques, S. A. Leonel, and P. Z. Coura,
Phys. Rev. B {\bf 72}, 094418 (2005).

\bibitem{benfatto_prb09}
L. Benfatto,  C. Castellani and T. Giamarchi,
Phys. Rev. B {\bf 80}, 214506 (2009)

\bibitem{meir_epl10}
A. Erez and Y. Meir, Europhys. Lett. {\bf 91}, 47003 (2010).

\bibitem{benfatto_review14}
L. Benfatto,  C. Castellani and T. Giamarchi, {\em Berezinskii-Kosterlitz-Thouless Transition within the Sine-Gordon Approach: The Role of the Vortex-Core Energy}, invited chapter for  {\em 40 Years of Berezinskii-Kosterlitz-Thouless Theory}, edited by Jorge V. Jos\`e  (World Scientific, Singapore, 2013).



\bibitem{mirlin_prb15}
E. J. K\:onig, A. Levchenko, I. V. Protopopov, I. V. Gornyi, I. S. Burmistrov, and A. D. Mirlin,
Phys. Rev. B {\bf 92}, 214503 (2015)


\bibitem{meir_prl13}
A. Erez and Y. Meir, \prl\ {\bf 111}, 187002 (2013).

\bibitem{maccari_prb17}
I. Maccari, L. Benfatto, and C. Castellani, Phys. Rev. B {\bf 96}, 060508 (R) 2017

\bibitem{maccari_cm18}
I. Maccari, L. Benfatto, and C. Castellani, Condens. Matter {\bf 3}(1), 8 (2018)

\bibitem{sacepe_11}
B.~Sac\'ep\'e, C. Chapelier, T. I. Baturina, V. M. Vinokur, M. R. Baklanov, M. Sanquer, Nature Communications {\bf 1}, 140 (2010). B.~Sac\'ep\'e {\em et al.}, Nature Phys. {\bf 7},  239 (2011).

\bibitem{mondal_prl11}
M.~Mondal, A.~Kamlapure, M.~Chand, G.~Saraswat, S.~Kumar,
J.~Jesudasan, L. Benfatto, V.~Tripathi, and P.~Raychaudhuri,  \prl\ {\bf 106}, 047001 (2011).


\bibitem{pratap_13}
A. Kamlapure, T. Das, S. Chandra Ganguli, J. B. Parmar,
S. Bhattacharyya, and P. Raychaudhuri, Sci. Rep. {\bf 3}, 2979 (2013).

\bibitem{noat_prb13}
Y. Noat, V. Cherkez,C. Brun,T. Cren,  C. Carbillet, F. Debontridder,  K. Ilin, M. Siegel, A. Semenov, H.-W. H\:ubers, D. Roditchev, \prb\  {\bf 88}, 014503 (2013).

\bibitem{roditchev_natphys14}
 C. Brun,	T. Cren,	V. Cherkez,	F. Debontridder,	S. Pons,	D. Fokin,	M. C. Tringides,	S. Bozhko, L. B. Ioffe, B. L. Altshuler	and D. Roditchev, Nat. Phys. {\bf 10}, 444 (2014). 

\bibitem{leridon_prb16}
C. Carbillet, S. Caprara, M. Grilli, C. Brun, T. Cren, F. Debontridder, B. Vignolle, W. Tabis, D. Demaille, L. Largeau, K. Ilin, M. Siegel, D. Roditchev, and B. Leridon, Phys. Rev. B {\bf 93}, 144509 (2016).

\bibitem{brun_review17}
C. Brun, T. Cren and D. Roditchev, Supercond. Sci. Technol. {\bf 30}, 013003 (2017)


\bibitem{trivedi_prb01}
A. Ghosal, M. Randeria, and N. Trivedi, Phys. Rev. B 65,
014501 (2001).

\bibitem{dubi_nat07}
Y. Dubi, Y. Meir and Y. Avishai, Nature  {\bf 449}, 876 (2007).

\bibitem{ioffe}
L. B. Ioffe and M. Mezard, Phys. Rev. Lett. 105, 037001 (2010);
M.V. Feigelman, L. B. Ioffe, and M.Mezard, Phys. Rev. B 82,
184534 (2010).

\bibitem{nandini_natphys11}
K. Bouadim, Y. L. Loh, M. Randeria, and N. Trivedi, Nat. Phys. {\bf 7}, 884 (2011).

\bibitem{seibold_prl12}
G. Seibold, L. Benfatto, C. Castellani, J. Lorenzana, 
Phys. Rev. Lett. {\bf 108}, 207004 (2012).

\bibitem{lemarie_prb13}
G. Lemari\'e, A. Kamlapure, D. Bucheli, L. Benfatto,
J. Lorenzana, G. Seibold, S. C. Ganguli, P. Raychaudhuri, and
C. Castellani, Phys. Rev. B {\bf 87}, 184509 (2013).


\bibitem{biscaras_natmat13}
J.  Biscaras,  N.  Bergeal,  S.  Hurand,  C.  Feuillet-Palma,
A. Rastogi, R. C. Budhani, M. Grilli, S. Caprara, and J. Lesueur,
Nat. Mater. {\bf 12}, 542 (2013).

\bibitem{bid_prb16}
Gopi Nath Daptary, Shelender Kumar, Pramod Kumar, Anjana Dogra, N. Mohanta, A. Taraphder, and Aveek Bid,
Phys. Rev. B {\bf 94}, 085104 (2016).

\bibitem{jespersen_prb16}
G. E. D. K. Prawiroatmodjo, F. Trier, D. V. Christensen, Y. Chen, N. Pryds, and T. S. Jespersen
Phys. Rev. B {\bf 93}, 184504 (2016).

\bibitem{scopigno2016} N. Scopigno, D. Bucheli, S. Caprara, J. Biscaras, N. Bergeal, J. Lesueur, and M. Grilli,
Phys. Rev. Lett. {\bf 116}, 026804 (2016).

\bibitem{caviglia_natcomm18}
Holger Thierschmann, Emre Mulazimoglu, Nicola Manca, Srijit Goswami, Teun M. Klapwijk and
Andrea D. Caviglia, Nat. Comm. {\bf 9}, 2276 (2018).

\bibitem{hurand_prb19}
S. Hurand, A. Jouan, E. Lesne, G. Singh, C. Feuillet-Palma, M. Bibes, A. Barth\'el\'emy,
J. Lesueur, and N. Bergeal, \prb\ {\bf 99}, 104515 (2019).



\bibitem{caprara_prb11} S. Caprara, M. Grilli, L. Benfatto, C. Castellani, Phys. Rev. B \textbf{84}, 014514 (2011).

\bibitem{caprara_sust15} S. Caprara, D. Bucheli, N. Scopigno, N. Bergeal, J. Biscaras, S. Hurand, J. Lesueur and M. Grilli, Supercond. Sci. Technol. \textbf{28} 014002 (2015).

\bibitem{moler_natmat13}
Beena Kalisky, Eric M. Spanton, Hilary Noad, John R. Kirtley, Katja C. Nowack, Christopher Bell, Hiroki K. Sato, Masayuki Hosoda, Yanwu Xie, Yasuyuki Hikita, Carsten Woltmann, Georg Pfanzelt, Rainer Jany, Christoph Richter, Harold Y. Hwang, Jochen Mannhart and Kathryn A. Moler, 
Nat. Mater. {\bf 12}, 1091 (2013).

\bibitem{moler_prb16}
Hilary Noad, Eric M. Spanton, Katja C. Nowack, Hisashi Inoue, Minu Kim, Tyler A. Merz, Christopher Bell, Yasuyuki Hikita, Ruqing Xu, Wenjun Liu, Arturas Vailionis, Harold Y. Hwang, and Kathryn A. Moler, Phys. Rev. B {\bf 94}, 174516 (2016).


\bibitem{kalisky_natmat17}
Yiftach Frenkel, Noam Haham, Yishai Shperber, Christopher Bell, Yanwu Xie, Zhuoyu Chen, Yasuyuki Hikita, Harold Y. Hwang, Ekhard K. H. Salje and Beena Kalisky, 
Nat. Mater. {\bf 16}, 1203 (2017).

\bibitem{kalisky_prb17} 
Shai Wissberg and Beena Kalisky, 
Phys. Rev. B {\bf 95}, 144510 (2017).

\bibitem{caprara_prl12}
S. Caprara, F. Peronaci, and M. Grilli, 
Phys. Rev. Lett. {\bf 109}, 196401 (2012).


\bibitem{triscone_science07}
N. Reyren, S. Thiel, A. Caviglia, L. F. Kourkoutis, G. Hammerl,
C. Richter, C. Schneider, T. Kopp, A.-S. R\:uetschi, D. Jaccard,
M. Gabay, D. Muller, J.-M. Triscone, and J. Mannhart, Science
{\bf 317}, 1196 (2007).

\bibitem{han_apl14}
Y.-L. Han, S.-C. Shen, J.You, H.-O. Li, Z.-Z. Luo, C.-J. Li, G.-L.
Qu, C.-M. Xiong, R.-F. Dou, L. He, D. Naugle, G.-P. Guo, and
J. Nie, Appl. Phys. Lett. {\bf 105}, 192603 (2014).

\bibitem{caviglia_prb17}
A. M. R. V. L. Monteiro, D. J. Groenendijk, I. Groen, J. de Bruijckere, R. Gaudenzi, H. S. J. van der Zant, and A. D. Caviglia, \prb\ {\bf 96}, 020504(R) (2017).


\bibitem{pearl}
J. Pearl, Appl. Phys. Lett. \textbf{5}, 65 (1964).



\bibitem{pratap_prb17}
Indranil Roy, Prashant Chauhan, Harkirat Singh, Sanjeev Kumar, John Jesudasan, Pradnya Parab, Rajdeep Sensarma, Sangita Bose, and Pratap Raychaudhuri, Phys. Rev. B {\bf 95}, 054513 (2017).

\bibitem{ye_science15}
J. M. Lu, O. Zheliuk, I. Leermakers, N. F. Q. Yuan, U. Zeitler,  K. T. Law,  J. T. Ye, Science {\bf 350}, 1353 (2015)

\bibitem{castroneto_nat16}
L. J. Li, E. C. T. O'Farrell, K. P. Loh,  G. Eda, B. \:Ozyilmaz, and A. H. Castro Neto, Nature {\bf 529}, 185 (2016).

\bibitem{pasupathy_natphys16}
A. W. Tsen, B. Hunt, Y. D. Kim, Z. J. Yuan, S. Jia, R. J. Cava, J. Hone, P. Kim, C. R. Dean and A. N. Pasupathy, Nature Physics {\bf 12}, 208 (2016).

\bibitem{biscaras_natcomm10}
J. Biscaras, N. Bergeal, A. Kushwaha, T. Wolf, A. Rastogi, R.C. Budhani and J. Lesueur,
Nat. Comm. {\bf 1}, 89 (2010)


\bibitem{landauer}R. Landauer, in \emph{Electrical Transport and Optical Properties of Inhomogeneous Media}, edited by J. C. Garland and D. B. Tanner (American Institute of Physics, New York, 1978), p. 2.
\bibitem{kirkpatrick}S. Kirkpatrick, Rev. Mod. Phys. {\bf 45}, 574 (1973).




\bibitem{amb-bar} V. Ambegaokar and A. Baratoff, Phys. Rev. Lett. \textbf{10}, 486 (1962); erratum, \textbf{11}, 104 (1963).

\bibitem{review_likharev}
K. K. Likharev, Rev. Mod. Phys. {\bf 51}, 101 (1979).

\bibitem{dezi} G. Dezi, N. Scopigno, S. Caprara, and M. Grilli, Phys. Rev. B {\bf 98}, 214507 (2018).

\end{thebibliography}
\end{document}